\documentclass[12pt]{article}

\usepackage{graphicx}
\usepackage{amsmath,amssymb}
\usepackage{bm}
\allowdisplaybreaks
\begin{document}
\baselineskip=0.7cm

\newcommand{\e}{{\rm e}}
\renewcommand{\theequation}{\arabic{section}.\arabic{equation}}
\newcommand{\Tr}{{\rm Tr}}
\newcommand{\np}{{\rm :}}
\renewcommand{\figurename}{Fig.\@}
\renewcommand{\thesection}{\arabic{section}.}
\renewcommand{\thesubsection}{\arabic{section}.\arabic{subsection}}
\makeatletter
\def\section{\@startsection{section}{1}{\z@}{-3.5ex plus -1ex minus 
 -.2ex}{2.3ex plus .2ex}{\large}} 
\def\subsection{\@startsection{subsection}{2}{\z@}{-3.25ex plus -1ex minus 
 -.2ex}{1.5ex plus .2ex}{\normalsize\it}}
\def\appendix{
\par
\setcounter{section}{0}
\setcounter{subsection}{0}
\def\thesection{\Alph{section}}}

\makeatother
\def\thefootnote{\fnsymbol{footnote}}

\begin{flushright}
hep-th/0612262\\
UT-KOMABA/06-15\\
Brown-HET-1476 \\
December 2006
\end{flushright}
\vspace{0.3cm}

\begin{center}
{\Large  $1/2$-BPS Correlators as $c=1$ S-matrix}

\vspace{0.7cm}

\normalsize
 \vspace{0.4cm}
Antal {\sc Jevicki}
\footnote{e-mail address:\ \ {\tt antal@het.brown.edu}}

\vspace{0.3cm}
{\it Department of Physics, Brown University\\
Providence, RI 02912
}

\vspace{0.5cm}

Tamiaki {\sc  Yoneya}
\footnote{
e-mail address:\ \ {\tt tam@hep1.c.u-tokyo.ac.jp}}

\vspace{0.3cm}

{\it Institute of Physics, University of Tokyo\\
Komaba, Meguro-ku, Tokyo 153-8902}

\vspace{1cm}
Abstract
\end{center}

We argue from two complementary viewpoints of Holography 
that the 2-point correlation functions 
of 1/2-BPS multi-trace operators in the large-$N$ (planar) 
limit 
are nothing but  the (Wick-rotated) S-matrix elements 
of $c=1$ matrix model. 
On the bulk side, we consider an Euclideanized version 
of the so-called bubbling geometries and show that the  
corresponding droplets reach the conformal boundary. 
Then the scattering matrix 
of fluctuations of the droplets gives directly the two-point correlators through the GKPW prescription. 
On the Yang-Mills side, we show that the two-point 
correlators of holomorphic and anti-holomorphic 
operators are essentially equivalent with the 
transformation functions 
between asymptotic {\it in}- and {\it out}-states of 
$c=1$ matrix model.  Extension to non-planar case is also 
discussed. 

\newpage
\section{Introduction}
A certain class of correlation functions of 
${\cal N}=4$ super Yang-Mills theory can be 
computed exactly using a free-field approximation. 
A very interesting nontrivial example of such cases 
is the two-point correlator of 
multi-trace local operators belonging to the 
$1/2$-BPS sector. In this case, the free-field approximation of the 
gauge theory can further be mapped to a 
one-dimensional matrix model \cite{cjr,ber}.
Hence, such correlators can be conveniently 
classified by using the language of one-dimensional non-relativistic 
free fermions. It is quite remarkable that 
the similar fermion-liquid (`droplet') 
 picture also arises in the 
holographically dual description, namely, 
on the supergravity side \cite{llm} as the classification 
of a class of the so-called bubbling geometries with 
the same amount of (super)symmetries as the 
$1/2$-BPS sector of the gauge theory. The correspondence 
of both sides in this particular sector can be regarded as 
yet another evidence for the AdS/CFT correspondence. 

It would be of some  interest to see 
whether this correspondence 
of {\it classification} at the level of spectrum 
could be further extended to 
agreements of physical observables. 
The purpose of the present note is to provide such an example. 
According to the standard prescription of the AdS/CFT 
correspondence,  
the gauge-theory correlators are 
interpreted as the amplitudes of propagation of supergravity modes 
in the bulk, connecting conformal boundary to conformal boundary 
in the AdS background \cite{gkww}. 
However, it has  not been 
clear what should be the corresponding 
interpretation of the above $1/2$-BPS two-point correlators 
for general multi-trace operators. 

We will argue that 
a reasonable holographic 
interpretation of two-point correlators 
 naturally emerges if we consider 
the bubbling geometries in an Euclideanized 
picture. The two-dimensional droplets 
characterizing the bubbling geometries are then 
infinite domains which extend to the conformal boundary 
of the asymptotically EAdS background.\footnote{
This possibility was first emphasized in \cite{yone}. 
}
As opposed to a circular droplet which is 
located close to the center of geometries 
in the Lorentzian case, 
the ground-state  droplet is an infinite wedge
 region bounded by a hyperbola. 
This can be related to the fermi sea of the ground state of the  
$c=1$ matrix model. Since the excited states of the droplet 
propagate along the hyperbola, we can naturally 
identify the scattering amplitudes of excitation modes 
along the hyperbola to be the correlators on the 
gauge theory side through the standard interpretation. 

On the other hand, from 
the viewpoint of the $c=1$ matrix model, we can 
argue that the correlators of $1/2$-BPS multi-trace 
operators which take a form of correlation 
between holomorphic and anti-holomorphic matrix operators 
is reinterpreted as defining an 
S-matrix as the amplitudes of transformation between incoming and 
outgoing multi-particle states of excitations 
near the fermi sea. 

In the next section, we first describe an Euclideanized 
version of bubbling geometries. Then in section 3, we 
discuss the scattering amplitudes 
(which we call  `boundary S-matrix') of excited modes 
along the ground-state droplet 
in classical approximation. 
In section 4, we discuss some examples of matching between 
boundary S-matrix and the correlators. We also  
study the limit of large R-charge momentum. 
The latter will explain the origin 
of a previous partial proposal 
due to ref. \cite{okuyama}
for a possible relation of the $1/2$-BPS 
correlators with the vertex of collective field theory 
of the $c=1$ matrix model. 
In section 5, we discuss the S-matrix interpretation 
of $1/2$-BPS correlators from the standpoint 
of the $c=1$ matrix model {\it per se}. 
This allows us to provide an intrinsic  
interpretation of the correlators as an S-matrix 
within a logic of 
matrix model. We also provide a nontrivial example of 
higher-genus effect which shows the 
correspondence of $1/2$-BPS  correlators 
with the $c=1$ S-matrix to arbitrary genera 
in the large momentum limit.  

\section{Euclideanized bubbling geometries} 

It is useful to first recall how the ground state, 
AdS$_5\times $S$^5$ geometry, is embedded in the 
LLM metric \cite{llm} in the Lorentzian case, 
\begin{equation}
ds^2=-h^{-2}(d\tau+V_idx^i)^2 + h^2 (dy^2 +dx^idx^i)
+y\e^Gd\Omega_3^2+y\e^{-G}d\tilde{\Omega}_3^2.
\end{equation}
All of the functions in this metric (and also the RR-fields which 
are suppressed here) are determined
\footnote{
The equations are $
h^{-2}=2y\cosh G, \quad z={1\over 2}\tanh G, \quad 
y\partial_y V_i=\epsilon_{ij}\partial_j z, \quad 
y(\partial_iV_j-\partial_jV_i)=
\epsilon_{ij}\partial_y z. $
} by 
specifying the value, either $1/2$ or $-1/2$, of 
a scalar function $z(x_i, y)$ on the plane D at $y=0$.  
Under this boundary condition, the Laplace equation for $z(x_i, y)$ 
\begin{equation}
\partial_i\partial_i z+y\partial_y({\partial_y z\over y})=0
\end{equation}
is 
solved as 
\begin{equation}
z(x_1, x_2, y)=
{y^2\over \pi}\int_{{\rm D}}{dx'_1dx'_2}
{z(x_1',x_2', 0)\over 
[(x-x')^2 +y^2]^2}.
\end{equation}

The ground state corresponds to a circular droplet 
of radius $r_0=\sqrt{4\pi g_sN}$ 
\begin{equation}
z(x_1, x_2,0)=\left\{\begin{array}{ll}
-1/2 & \mbox{$(x_1, x_2) \in$ circular disk of radius $r_0$}\\
+1/2 & \mbox{otherwise}
\end{array}
\right. 
\end{equation}
which gives 
\begin{align}
z(x_1, x_2, y)-{1\over 2}
&=-{y^2\over \pi}\int_{{\rm Disk}}{r'dr'd\phi'\over 
r^2+r^{'2}-2rr'\cos\phi'+y^2]^2}\cr
&={r^2-r_0^2 +y^2\over 2
\sqrt{(r^2+r_0^2+y^2)^2-4r^2r_0^2}}-{1\over 2}, 
\end{align}
using spherical coordinates for the 2-dimensional 
plane $(x_1, x_2)=r(\cos \phi, \sin\phi)$. Then defining 
new coordinates $\rho, \theta$ and $\tilde{\phi}$  by
\begin{equation}
y=r_0\sinh \rho \sin\theta, \quad r=r_0\cosh\rho\cos\theta, \quad 
\tilde{\phi}=\phi-\tau, 
\end{equation}
the LLM metric above reduces to the standard 
AdS$_5\times $S$^5$ metric expressed in terms 
of the global coordinate, 
\begin{equation}
ds^2=r_0\Big[
-\cosh^2 \rho d\tau^2 +d\rho^2 +\sinh^2\rho d\Omega_3^2
+d\theta^2 +\cos^2\theta d\tilde{\phi}^2 + \sin^2\theta d\tilde{\Omega}_3^2
\Big].
\end{equation}

Let us now perform a (double) Wick rotation \cite{dsy} 
$\tau\rightarrow -i\tau, 
\phi\rightarrow -i\psi\, \, ( \tilde{\psi}\equiv
\psi-\tau \rightarrow -i\tilde{\psi})$ under which both 
the AdS metric and the RR-fields are transformed 
`covariantly' into the Euclideanized AdS (EAdS$_5\times$ S$^{4,1}$)
background with the metric, 
\begin{equation}
ds^2=r_0\Big[
\cosh^2 \rho d\tau^2 +d\rho^2 +\sinh^2\rho d\Omega_3^2
+d\theta^2 -\cos^2\theta d\tilde{\psi}^2 + \sin^2\theta d\tilde{\Omega}_3^2
\Big]. 
\end{equation}
Since the signature of this metric is still $9+1$ in 10-dimensional 
sense, supersymmetries can be preserved by a suitable 
renaming of spinor variables. 
The two-dimensional coordinates $(x_1,x_2)$ are then 
transformed as 
\begin{equation}
x_1\rightarrow x_1=r\cosh \psi, \quad 
x_2\rightarrow ix_2=i r\sinh \psi .
\label{hyperbola}
\end{equation}
This exercise implies 
that for discussing generic Euclideanized LLM ansatz, 
it is sufficient to make the double 
Wick rotations $x_2\rightarrow ix_2$ 
and $\tau\rightarrow -i\tau$. The vector field $V_i$ 
must also be rotated covariantly as 
$V_1\rightarrow -iV_1, \, \,  V_2\rightarrow V_2$. 

Therefore the Laplace equation now becomes a 
hyperbolic wave equation
\begin{equation}
(\partial_1^2-\partial_2^2)z+y\partial_y({\partial_y z\over y})=0. 
\end{equation}
Although it is actually sufficient to consider a 
wedge region $x_1^2-x_2^2\ge 0, \, x_1\ge 0$ which we 
hereafter denote by the symbol W, 
for discussing the 
asymptotically EAdS backgrounds, 
let us  consider this wave equation 
in the whole (Wick-rotated) $(x_1,x_2)$ plane which we denote 
by the same symbol D as in the Lorentzian bubbling geometries. 
The region outside the W will be denoted by $\overline{{\rm W}}\equiv$  
D$-$W. 
We have to specify the boundary condition 
at $y=0$ for the whole plane 
D just as in the Lorentzian case by demanding smoothness 
of solutions at $y=0$. 

The solution must satisfy the boundary condition 
$z(x_i, y)|_{y=0}=z(x_i, 0)=1/2$ or $-1/2$ at $y=0$. 
Under this boundary condition, 
the appropriate solution which includes the above
EAdS background as the ground state
is 
\begin{equation}
z(x_1, x_2, y)=-
{y^2\over \pi}{\rm Im}\Big[\int_{{\rm D}}{dx'_1dx'_2}
{z(x_1',x_2', 0)\over 
[(x_1-x'_1)^2-(x_2-x_2')^2 +y^2-i\epsilon]^2}\Big]. 
\label{bsolution}
\end{equation}
This is verified by noting that 
\[
\lim_{y\rightarrow 0}\Big[{y^2\over 
[(x_1-x'_1)^2-(x_2-x_2')^2 +y^2 - i\epsilon]^2}-
{y^2\over 
[(x_1-x'_1)^2-(x_2-x_2')^2 +y^2 + i\epsilon]^2}\Big]
\]
\begin{equation}
=-2 \pi i \delta(x_1-x_1')\delta(x_2-x_2').
\end{equation}
We here chose 
 the boundary condition outside the wedge region  W as
\begin{equation}
z(x_1', x_2', 0) = 1/2  \quad \mbox{for} \quad (x_1',x_2') \in \overline{{\rm W}}. 
\end{equation}
After obtaining solutions, we can restrict ourselves to 
the wedge region $(x_1, x_2) \in$ W for general nonzero 
$y$, which is related to the Euclidean 
plane D of Lorentzian solutions by \eqref{hyperbola}. 

For example, the EAdS solution discussed above is 
obtained by replacing  
the circular disk (with value $-1/2$) of the Lorentzian case 
by an infinite domain H ($0< r<r_0$) in the wedge region 
W bounded by a hyperbola at $r=r_0$. 
It can be checked by explicitly evaluating the 
above integral that the expression 
of $z$ for the EAdS takes the same form as in the Lorentzian case 
\begin{align}
&-{y^2\over \pi}{\rm Im}\Big[\int_{{\rm D}}{dx'_1dx'_2}
{z(x_1',x_2', 0)\over 
[(x_1-x'_1)^2-(x_2-x_2')^2 +y^2-i\epsilon]^2}\Big]\cr
&=1/2 +{y^2\over \pi}{\rm Im}\Big[\int_{{\rm H}}{dx'_1dx'_2}
{1\over 
[(x_1-x'_1)^2-(x_2-x_2')^2 +y^2-i\epsilon]^2}\Big]\cr
&={1\over 2}{r^2-r_0^2 +y^2\over 
\sqrt{(r^2+r_0^2+y^2)^2 -4r^2r_0^2}} \rightarrow 
\left\{ 
\begin{array}{ll} 1/2 & \mbox{if $r>r_0\, ,  \, y=0$} \\
-1/2 & \mbox{if $r<r_0 \, ,  \, y=0$}
\end{array}
\right.
\end{align}
as it should be, since the ground state droplet in either metric 
has no explicit dependence on the angle $\phi\leftrightarrow i\psi$. 
 
Thus we can define Euclideanized bubbling geometries 
by setting hyperbolical droplet at $y=0$ in a similar 
 way as in the Lorentzian case.  
The excited droplets can have arbitrary holes and 
islands of droplets in the wedge region W. 
The hyperbolical shape of the ground-state droplet implies that 
in comparing with the free-fermion picture 
of matrix models, we should consider the usual 
$c=1$ model  with inverted harmonic oscillator potential, 
instead of the ordinary harmonic potential of the Lorentzian case. 
There are two regions (corresponding to 
positive or negative values of $x_1$) separated 
by the inverted potential. 
But for discussing deformations of the ground-state 
hyperbola in the classical approximation, it is sufficient 
to consider only one of them. 
 Since after all we are discussing the classical supergravity solutions 
which are only justified in the large $N$ limit, 
we are allowed to treat the fermi sea in the semi-classical 
approximation on the matrix-model side too. 

\begin{figure}[htbp]
\begin{center}
\begin{minipage}[t]{0.4\textwidth}
\begin{center}
 \includegraphics[width=3cm,clip]{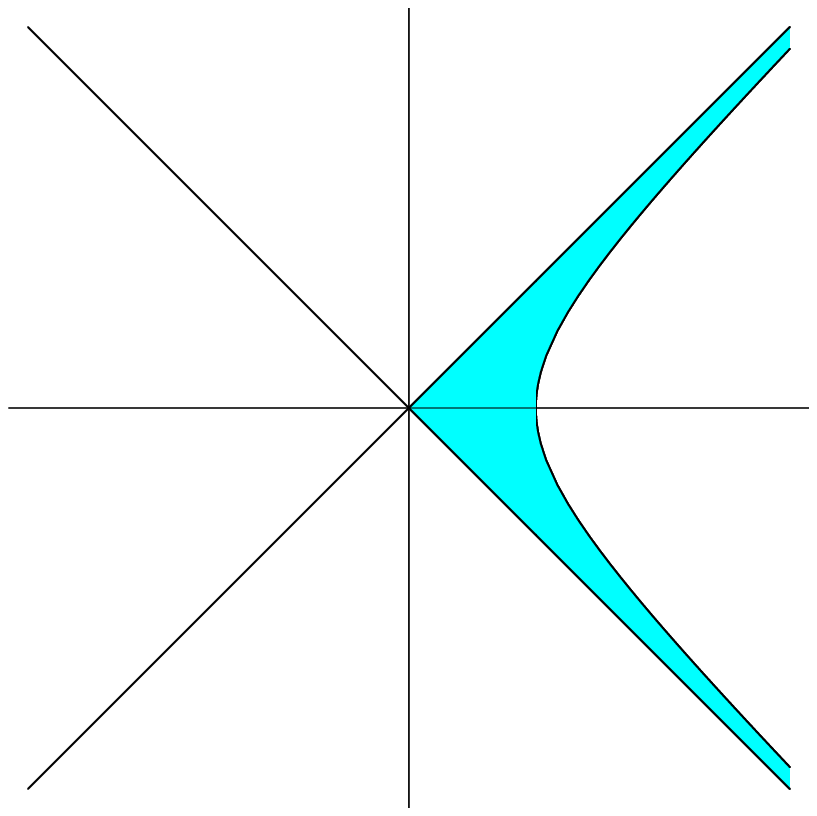}
\caption{{\footnotesize The shaded region (H) in the wedge region 
W in the plane D represents 
the ground-state droplet. }}
\end{center}
\end{minipage}
\hspace{1cm}
\begin{minipage}[t]{0.4\textwidth}
\begin{center}
 \includegraphics[width=3cm,clip]{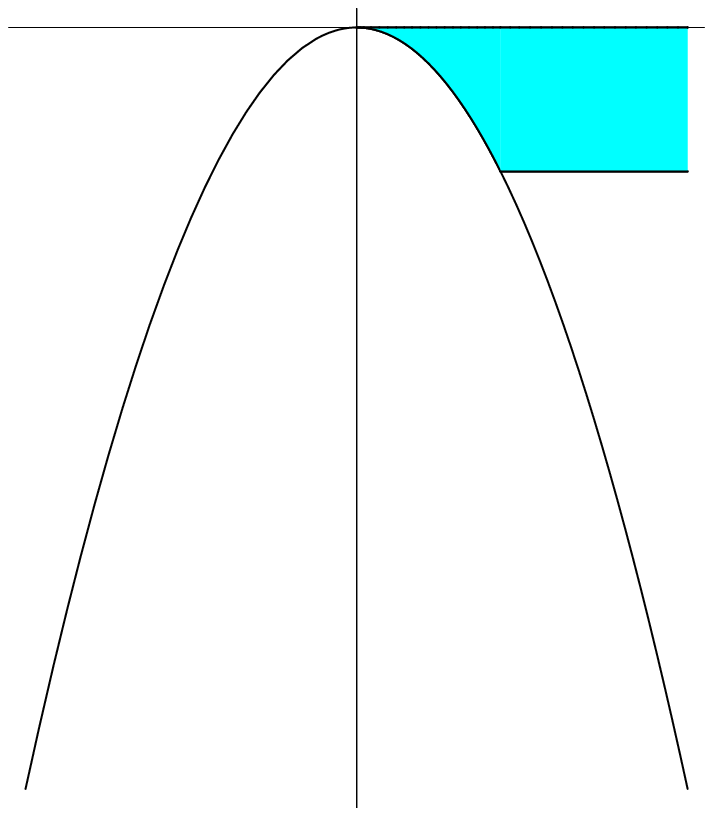}
\caption{{\footnotesize  In the fermion picture, 
the  ground state droplet corresponds to filling 
levels from the top of the potential. }}
\end{center}
\end{minipage}\end{center}
\end{figure}

There is actually a subtle difference 
that the droplet should be 
considered only in the wedge region W. In terms of fermionic 
picture, this amounts to considering one-particle 
levels only 
up to the top of the inverted harmonic potential. See Figs.1 and 2. 
The levels above the potential must be totally 
ignored, and the ground state is the one in which the 
levels are filled from the top of potential, {\it not} 
from the (infinite) bottom of  a suitably regularized  
potential as usual. In other words, 
particles and holes are interchanged there, comparing with 
 case of the usual $c=1$ matrix model.  Clearly, however, 
these differences do not matter again in the semi-classical 
treatment of droplets.

Now going back to the EAdS geometry obtained by 
our Wick rotation, the wedge region W of the droplet plane 
reaches the conformal boundary 
corresponding to the limit $\psi\rightarrow \pm\infty$ 
which implies 
$\rho \rightarrow \infty$. In discussing the boundary-to-boundary 
 propagation of supergravity modes in 
the AdS background, it is usually  
most convenient to use Poincar\'{e} coordinates. 
Therefore let us try to define the corresponding coordinates 
for general (Euclidean) LLM geometries. Since we will be 
interested in deformations of the ground-state droplet 
in the fixed plane W with two SO$_4$ symmetries kept intact, 
it is sufficient to deal with four coordinates  $(\tau, x_1, x_2, y)$. 
They can 
be reparametrized as $(\tau, \rho, \psi, \theta)$ by 
defining 
\begin{equation}
x_1=r\cosh \psi, \quad x_2=r\sinh \psi, \quad r=
r_0\cosh \rho \cos \theta, \quad 
y=r_0\sinh \rho \sin \theta 
\label{poincare}
\end{equation}
which are the same coordinates as we have used for the EAdS 
background. 
Then, for the two-dimensional part described by $(\tau, \rho)$, we can make 
a  coordinate transformation to $(u, v_4)$ defined below, where 
the direction of $v_4$ is interpreted as 
the 4-th direction of the conformal boundary, by 
assuming that we restrict ourselves to the bubbling 
geometries which reduce to EAdS background asymptotically 
as $u\rightarrow \infty$:
\begin{align}
\cosh\rho \cosh \tau&={1\over 2u}\Big(1+u^2(1+v_4^2)\Big),\\
\cosh\rho\sinh \tau&=uv_4,\\
\sinh \rho&={1\over 2u}\Big(1-u^2(1-v_4^2)\Big).
\end{align}
In the case of the EAdS, this two-dimensional 
part reduces to the two-dimensional section $(u, v_4)$ of the 
Poincar\'{e} patch $(u, v_1,v_2,v_3,v_4)$, 
whose metric is nothing but the 
EAdS$_2$,  $u^{-2}du^2 +u^{2}dv_4^2$.  
as is easily seen by comparing this 
definition with the ordinary Poincar\'{e} coordinates 
of the EAdS$_5$ background. 

Eqs. \eqref{poincare} show that in terms of the coordinates $\rho$ and $\theta$, the 
droplet plane D at $y=0$ is described by two patches, either 
$\rho=0$ or $\theta=0$ as in the  case of Lorentzian convention. 
Inside the region W, 
the former corresponds to $0\le x_1^2-x_2^2\le r_0^2$, 
while the later to $x_1^2-x_2^2\ge r_0^2$. 
The EAdS ground state corresponds to the droplet 
where the first patch is completely filled.  The 
droplets of general excited states have holes inside 
the first patch, and the second patch has nonzero occupied 
regions. If we take into account the Euclidean 
time coordinate $\tau$, the  boundary  along which the 
two patches are sewn is the product space of 
hyperbola $x_1^2-x_2^2=r_0^2$ and a semi-circle defined by 
\begin{equation}
{1\over u^2}+ v_4^2=1 \quad 
(u=\cosh \tau, \, v_4=\tanh \tau) .
\end{equation}
The latter equation shows that the two-dimensional section 
in consideration 
reaches the conformal boundary $u\rightarrow \infty$ 
at $v_4=\pm 1$. If we wish, this can be regarded again as 
a hyperbola by defining new 
variables 
$
q=u, \, \,  p=uv_4
$
in terms of which the above equation is 
expressed as 
$
q^2-p^2=1. 
$
We note that the  semicircle coincides with the 
`tunneling trajectory' \cite{dsy, dy} which describes 
two-point functions of 
the BMN operators semi-classically using the GKPW relation. 
The time  $\tau$ plays the role 
of affine parameter along the trajectory. 

\section{Scattering of droplets}
\setcounter{equation}{0}
\subsection{Boundary S-matrix}
Consider first a form of droplets which can be represented by 
the profile of their boundary in W
\begin{equation}
x_1=a(\psi)\cosh \psi, \quad x_2=a(\psi)\sinh \psi, 
\end{equation}
\begin{equation}
a(\psi)=r_0+\tilde{a}(\psi). 
\end{equation}
The function $\tilde{a}(\psi)$ measures the 
deformation of the boundary profile of the ground-state 
droplet. In terms of the 
coordinate $\tilde{\psi}=\psi-\tau$ 
of the EAdS background,   
the above droplet should be reinterpreted as 
a $\tau$-dependent form 
\begin{equation}
x_1=a(\tilde{\psi})\cosh (\tau +
\tilde{\psi}), \quad x_2=a(\tilde{\psi})\sinh (\tau+ 
\tilde{\psi}). 
\end{equation} 
In what follows, we rename $\tilde{\psi}$ by $\psi$ for 
notational simplicity. 
The rationale for this interpretation of  time dependent droplet 
 is that  the equation of small deformations 
around the EAdS background must be 
\begin{equation}
{d^2 x_i\over d\tau^2}=x_i, \label{eqmotion0}
\end{equation}
which is obtained by our Wick rotation from the 
corresponding equation of AdS background as 
established in ref. 
\cite{maoz} 
by studying the 
quantization of fluctuations around the AdS geometry. 
The ground state profile does not change under this 
motion, while the 
excited state profiles moves along the 
hyperbola, as is familiar in the treatment
of collective motions of fermi liquid
 in the inverted harmonic oscillator potential 
in the semi-classical approximation of the 
$c=1$ matrix model. 
The equation of supergravity 
around the EAdS background is obeyed by 
solutions obtained from these time dependent droplets at $y=0$ 
through \eqref{bsolution}.  

This implies  that the scattering amplitude of the excited 
states of Euclideanized bubbling geometries from conformal boundary to conformal boundary in the bulk 
essentially coincides with that of the deformed droplets propagating 
from $\tau=-\infty$ to $\tau=+\infty$. 
This is also consistent with 
the fact that  the energy of the droplet in supergravity
 is directly given \cite{llm} by the 
collective Hamiltonian of the droplet, together with the known 
observation that the symplectic structure 
of supergravity action also reduces \cite{maoz} 
to that of the 
collective fields of the droplet. 

Then, according to a general idea of the holographic 
correspondence, the GKPW relation, 
 between bulk-boundary propagators and 
correlators on the boundary, we naturally expect that 
the S-matrix of droplets will be directly 
connected to the two-point functions 
of multi-trace 1/2-BPS operators of the (Euclideanized)
 SYM, under an appropriate mapping between the deformations  
of the ground-state profile and the set of those local operators. 

There is a well-known method \cite{pol} for deriving the S-matrix of droplet in this picture which has been established in the context of 
$c=1$ matrix model. To make the present paper 
reasonably self-contained, we briefly review 
the basic idea. 
Renaming $x_i$ by $x_1\rightarrow x, \, x_2\rightarrow p$, 
the equation of motion for the profile in the phase space 
$(x, p_{\pm}(x, \tau))$ is 
\begin{equation}
\frac{\partial}{\partial \tau}p_{\pm}=x-p_{\pm}\frac{\partial}{\partial x}
p_{\pm}
\label{eqmotion1}
\end{equation}
where we interpret the momentum as a function of 
$(x, \tau)$ and the suffix $+$ and $-$ indicate 
two regions $p_+>0$  and $p_-<0$, respectively:
$p_{\pm}=\pm \sqrt{x^2-a^2}$. 
The S-matrix is by definition given by considering the relation 
between two asymptotic regions $\tau \rightarrow \pm \infty$. 

Following \cite{pol}, we set for sufficiently large $x$ 
\begin{equation}
x=\e^{q}  , \quad
p_{\pm}
=\pm \e^{q}\mp \e^{-q}\epsilon_{\pm}(q, \tau).
\end{equation}
Since in these asymptotic regions we have 
$
x\sim {a(\psi)\over 2}\e^{-\tau-\psi}
$ and $
\sim {a(\psi)\over 2}\e^{\tau+\psi}
$ for $\tau\rightarrow \mp \infty$ respectively, 
the $\epsilon$ fields behave, with respect to the 
dependence on $\tau$, as 
$\epsilon_{\pm}(q, \tau) \sim \epsilon_{\pm}(\tau\mp q)$. 
Hence the lapse of time between incoming ($\tau_i$) 
and outgoing ($\tau_i$)  
waves at a fixed large value of $x=\e^q$  
satisfies
\begin{equation}
\tau_f-\tau_i=2q+\log {a(\psi)^2\over 4},
\quad 
\epsilon_-(q, \tau_i)=\epsilon_+(q, \tau_f)={a(\psi)^2\over 2}
\end{equation}
which lead to a functional equation for asymptotic forms 
\begin{equation}
\epsilon_+(\tau-q)=\epsilon_-(\tau-q -
\log {\epsilon_+(\tau -q)\over 2}). 
\end{equation}
Or, using the fluctuating fields $\delta_{\pm}$ 
defined by 
$
{\epsilon_{\pm}(x)\over 2}=c_0^2+ \delta_{\pm}(x),
$
with $c_0$ being the scale of the static ground-state droplet, 
\begin{equation}
\delta_+(x)=\delta_-(x-\log (c_0^2 + \delta_+(x)). 
\end{equation}
The solution is given as  (see \cite{moore} for more details)
\begin{equation}
\delta_{\pm}(x)=
-c_0^2 \sum_{p=1}^{\infty}
\frac{(-1)^p\Gamma(\pm \partial_x +p-1)}{p!\Gamma(\pm \partial_x)}
\Big(\frac{\delta_{\mp}(x)}{c_0^2}\Big)^p. 
\end{equation}

This can be interpreted as the relation between 
$in$- and $out$-fields of collective fields $\chi_{\pm}$, 
defined by 
\begin{equation}
\delta_{\pm}\sim \sqrt{4\pi}(\partial_{\tau}\mp
\partial_q)\chi_{\pm}, \quad 
\chi_{\pm}(\tau\mp q)=-{i\over \sqrt{4\pi}}
\int \alpha_{\pm}(\xi)\e^{i\xi(\tau \mp q)}
\frac{d\xi}{\xi}
\end{equation}
satisfying the canonical commutation relations, 
\begin{equation}
[\alpha_{\pm}(\xi), \alpha_{\pm}(\xi')]=-\xi \delta(\xi'+\xi), 
\quad \alpha_{\pm}(-\omega)|0\rangle=0  \quad \omega>0. 
\end{equation}
In terms of the normal-mode operators in the momentum 
representation, we have 
\begin{equation}
\alpha_{\pm}(\eta)
=\sum_{p=1}^{\infty}
\Big({2\over c_0^2}\Big)^{p-1}
\frac{\Gamma(1\mp i\eta)}{\Gamma(2\mp i\eta -p)}{1\over p!}
\int d^p\xi\,  \delta(\eta-\sum \xi_i)
\Big(\prod \alpha_{\mp}(\xi_i)\Big)
\label{inoutrelation}
\end{equation}
and the S-matrix in the classical (tree) approximation 
is given by 
\begin{equation}
S(\sum \omega_i \rightarrow \sum \omega_i')
=\langle 0|\prod_i \alpha_-(-\omega_i')
\prod_j\alpha_+(\omega_j) |0\rangle. 
\end{equation}

For example, for $n\rightarrow 1$ scattering ($n\ge 2$),
 the S-matrix elements are, up to the delta-function of 
energy conservation ($\omega=\omega_1+\cdots \omega_n)$ 
which we will always suppress in what follows, 
\begin{equation}
\langle 0|\alpha_-(-\omega)\alpha_+(\omega_1)
\cdots \alpha_+(\omega_n)
|0\rangle 
\Rightarrow \Big({2\over c_0^2}\Big)^{n-1}(-i\omega)
(-i\omega-1)\cdots(-i\omega-n+2)\, 
\omega_1\cdots \omega_n .
\label{n1amplitude}
\end{equation}
In applying these results to our case, we have to 
Wick-rotate the momentum (=$\pm$energy on the mass-shell) as 
$\xi=\omega \rightarrow iJ$ with $J$ being the R-charge, 
in order to take into account
 our prescription of Euclideanized bubbling geometries. 
 Note that this procedure effectively makes the 
 {\it coupling constant}, $1/c_0^2$,  pure imaginary, 
 apart from an overall factor $i$ which can be absorbed into 
 Wick rotation of the 
 $\delta$-function of energy conservation.

\subsection{Hamiltonian formalism}
We finally recall that the same S-matrix elements as above 
are obtained 
by the Hamiltonian formalism of collective field theory \cite{je}. 
The equation \eqref{eqmotion1} can be recast in the 
Hamiltonian form 
\begin{equation}
\partial_{\tau}p_{\pm}=i[H, p_{\pm}]
\label{eqmotion2}
\end{equation}
by defining the effective Hamiltonian and the 
commutation relations as 
\begin{equation}
H=\int dx \Big[
\Big(\frac{p_+^3}{6}-\frac{(x^2+\mu)p_+}{2}\Big)
-
\Big(\frac{p_-^3}{6}-\frac{(x^2+\mu)p_-}{2}\Big)\Big], 
\end{equation}
\begin{equation}
[p_{\pm}(x, \tau), p_{\pm}(x', \tau)]=\mp i\delta'(x-x').
\end{equation}
Here, $\mu$ is an arbitrary constant, corresponding to an 
integration constant for the equation of motion. 
Note also that we use the usual Lorentzian convention 
in writing down the equations of motion, 
by interpreting  \eqref{eqmotion0} as 
being due to the inverted harmonic potential $V(x)=-x^2/2$. 
Actually, there is an ambiguity whether we take positive 
$\mu$ or negative one. Here we choose the positive convention 
for definiteness. 
This choice has an advantage in that the interaction 
Hamiltonian is not singular.  The negative $\mu$, 
which looks more natural in view of the 
form of the profile function,   would give a 
singular interaction Hamiltonian. 
Remarkably, however, there exists a duality
 that {\it both} give the 
{\it same} S-matrix 
with suitable regularization for the negative choice. 

To make the system look more 
 like a usual canonical system, 
 we define the shifted fields $\tilde{\phi}_{\pm}$ 
 (the sign of $\mu$ becomes relevant here): 
 \begin{equation}
p_{\pm}(x, \tau)=\pm (\sqrt{x^2+\mu}+ 
\tilde{\phi}_{\pm}(x,\tau)), 
\end{equation} 
\begin{equation}
H=\int dx \Big[
{1\over 2}\sqrt{x^2+\mu}(\tilde{\phi}_+^2+\tilde{\phi}_-^2)
+{1\over 6}(\tilde{\phi}_+^3 + \tilde{\phi}_-^3)
\Big], 
\end{equation}
which reduces,  by further making 
a change of variables 
$
x=\mu \sinh \sigma, \, 
\tilde{\phi}_{\pm}=\Big|{d\sigma\over dx}\Big|\phi_{\pm}, 
$
to
\begin{equation}
H=\int_0^{\infty}d\sigma \, \Big[
{1\over 2}(\phi_+^2+\phi_-^2)
+{1\over 6}\Big|{d\sigma\over dx}\Big|^2
(\phi_+^3 +\phi_-^3)
\Big].
\end{equation}
Here we have subtracted a (field-independent) c-number contribution. 
Then, using  the normal-mode expansion in the 
interaction representation, 
\begin{equation}
\phi_{\pm}(\sigma, \tau)=
{1\over \sqrt{2\pi}}\int_{-\infty}^{\infty}d\xi
\e^{-i\xi(\tau\mp \sigma)}\alpha(\xi)
\end{equation}
with 
$
[\alpha(\xi), \alpha(\xi')]
=-\omega\delta(\xi+\xi')
$
which can be identified with the normal-mode 
operators of $\chi_{\pm}$ introduced above 
in the asymptotic region $|\sigma|\sim q \rightarrow 
\infty$, we obtain 
\begin{equation}
H=H_2+ H_3(\tau), 
\label{collectivehamiltonian}
\end{equation}
\begin{equation}
H_2=\int_{0}^{\infty}d\xi \, 
\alpha(\xi)\alpha(-\xi), 
\end{equation}
\begin{equation}
H_3(\tau)=
{1\over 6}
\int_{-\infty}^{\infty} d^3\xi\, f(\xi_1+\xi_2+\xi_3)\, 
\e^{-i(\xi_1+\xi_2+\xi_3)\tau}
\alpha(\xi_1)\alpha(\xi_2)\alpha(\xi_3)
\label{h3}
\end{equation}
with
\begin{equation}
f(\xi)=\int_{-\infty}^{\infty} d\sigma
\, {1\over \mu \cosh^2 \sigma}\e^{i\xi\sigma}=
{2\pi \xi\over \mu\sinh \pi \xi}.
\end{equation}
  Note that 
the range of the $\sigma$-integral becomes  
the whole real axis by combining $\phi_{\pm}$ contributions
into a single integral, even though the original range 
was the half real axis $0< \sigma <\infty$. If we choose 
negative $\mu$, the form factor in $H_3$ would have been 
$1/\sinh^2 \sigma$ instead of $1/\cosh^2\sigma$.  
Using this interaction Hamiltonian, it is straightforward to 
compute the S-matrix in perturbative expansion in 
$1/\mu\propto  2/c_0^2$. The agreement of the results 
with the 
polynomial form exhibited in \eqref{inoutrelation} and 
\eqref{n1amplitude} 
has been confirmed in \cite{djr}\cite{kleb}\cite{jul}, 
together with some extensions to 
quantum corrections. This is somewhat miraculous 
in view of the presence of a nontrivial form factor $f(\xi)$ 
in the interaction Hamiltonian \eqref{h3}. 

\section{Comparison with the two-point functions of multi-trace 
operators}
\setcounter{equation}{0}
Let us recall the structure of a representative set of 
the chiral primary 1/2-BPS 
operators on the Yang-Mills side, which are 
characterized SO(4) symmetry and the conformal dimensions 
$\Delta =J$, 
\begin{equation}
 {\cal O}^J_{(J_1, J_2, \ldots, J_n)}(x)\equiv 
\Tr\Big(Z(x)^{J_1}\Big)\Tr\Big(Z(x)^{J_2}\Big)
\cdots \Tr\Big(Z(x)^{J_n}\Big)
, \quad J=\sum_i J_i
\label{halfbpsope}
\end{equation}
where $Z=(\phi_5+i\phi_6)/\sqrt{2}$ is the complex scalar 
field with a unit R-charge $J=1$ with respect to the rotation in the $5$-$6$ plane. 
Due to the non-renormalization property, two-point 
correlation functions of these operators with their conjugate 
set of operators constructed in terms of $\overline{Z} 
=(\phi_5-i\phi_6)/\sqrt{2}$ are given by 
the free-field results, 
\begin{equation}
\langle \overline{{\cal O}}^J_{(J'_1, J'_2, \ldots, J'_m)}(x')
\, {\cal O}^J_{(J_1, J_2, \ldots, J_n)}(x)\rangle
=F(\{(J'), (J)\}, N)D_4(x,x')^J
\end{equation}
where $D_4(x,x')\propto |x-x'|^{-2J}$ with 
$J=\sum_iJ_i=\sum_iJ'_i$ is the massless 
free-field propagator in $4$ dimensions and 
the function $F(\{(J'), (J)\}, N)$ is determined by 
the free-field contraction among the indices of scalar fields 
$\phi_i$ between $O^J(x)$ and $\overline{O}^J(x')$. 
Obviously, the  function $F$ is completely independent of 
spacetime coordinates. Furthermore, it is independent 
of the spacetime dimensions and signature. 
The double Wick-rotation we have discussed in the previous section 
requires us to Wick-rotate the angle coordinate in the 
$5$-$6$ plane. In terms of the above scalar fields, 
it amounts to rotating $\phi_6$ to pure imaginary axis,  
$i\phi_6$, and hence replace $Z$ ($\overline{Z}$) 
by $Z=(\phi_5-\phi_6)/\sqrt{2}$ \, ($\overline{Z}
=(\phi_5+\phi_6)/\sqrt{2}$) with $\phi_6$ now being 
quantized with {\it negative} metric. The crucial 
property $\langle Z\, Z\rangle =
\langle \overline{Z}\, \overline{Z}\rangle =0$ is 
preserved under this procedure, and hence this rotation 
does not affect the above final form of the 
correlators. 

For the specific purpose of the present section, 
comparison of the S-matrix elements with the correlators, 
it is sufficient to study only the function $F(\{(J'), (J)\}, N)$ which is 
independent of possible choices of  matrix models.  
Let us therefore suppress   the spacetime coordinates
and, hence,   spacetime dependent 
factor such as $D_4(x, x')$ in what follows. 
The simplest case $m=n=1$, we have
\begin{equation}
\langle 
\Tr \overline{Z}^J \Tr Z^J \rangle =JN^J. 
\end{equation}
The results of the previous section suggests that this trivial two-point function should be interpreted as 
the trivial $(1\rightarrow 1)$ S-matrix element 
\begin{equation}
\langle 0|\alpha_-(-iJ)\alpha_+(iJ)|0\rangle =J . 
\label{euclidnormal}
\end{equation}
As before  we have suppressed  an overall $\delta$-function factor $-i\delta(J-J)=-i\delta(0)$ of 
momentum conservation, which is common for all S-matrix elements. Then, the natural normalization 
for making correspondence between the 
S-matrix elements of the droplet and the correlators is 
$
\alpha(iJ) \leftrightarrow \frac{1}{N^{J/2}}\Tr 
Z^J
$ for incoming states and $
\alpha(-iJ) \leftrightarrow \frac{1}{N^{J/2}}\Tr 
\overline{Z}^J $ for outgoing states. 

Once single-trace operators are related to 
single-particle states, the multi-trace operators must 
be interpreted as multi-particle states. 
Under this interpretation, let us study some 
examples of ($n\rightarrow 1$) correlators in the leading 
planar approximation in the large $N$ limit. 
We present the results of graphical computations  
for $n=2,3,4$ ($J=\sum_{i=1}^nJ_i$). 
\begin{align}
&F(J, \{J_1, J_2\}, N)_{planar}
=
 JJ_1J_2 N^{-1}, 
\\
&F(J, \{J_1, J_2,J_3\}, N)_{planar} 
=J\Big(
J_1J_2J_3(J_1-1)+J_1J_2J_3(J_2-1)\nonumber \\
&+J_1J_2J_3(J_3-1) + 2J_1J_2J_3
\Big)N^{-2}
=J_1J_2J_3J(J-1) N^{-2}
\end{align}
where for $n=3$ 
the first 3 contributions come 
from `chain'-type diagrams 
in which the 3 traces of the initial state connected 
like a chain of 3 beads, while the last one comes from 
a `clover'-type diagram in which all of the 3 traces are 
connected simultaneously at a single point. 
\begin{figure}[htbp]
\begin{center}
\begin{minipage}[t]{0.88\textwidth}
\begin{center}
 \includegraphics[width=10cm,clip]{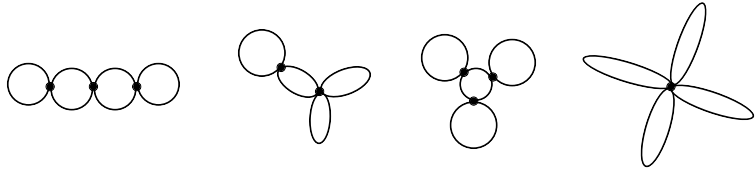}
\caption{{\footnotesize  Four types of planar contractions of 
multiple traces into a single trace for $1\leftrightarrow 4$ 
scattering.  }}
\end{center}
\end{minipage}
\end{center}
\end{figure}
\vspace{-1cm}
\begin{align}
F(J, \{J_1, J_2, J_3, J_4\}, N)_{planar}&=J\Big[
2\Big(J_1J_2(J_2-1)J_3(J_3-1)J_4 \nonumber \\
&+J_1J_3(J_3-1)J_4(J_4-1)J_2 +J_1J_2(J_2-1)J_4(J_4-1)J_3\nonumber \\
&+J_2J_1(J_1-1)J_4(J_4-1)J_3 + 
J_3J_1(J_1-1)J_2(J_2-1)J_4 \nonumber \\
&+ J_2J_1(J_1-1)J_3(J_3-1)J_4\Big)
\nonumber\\
&+
6\Big(
J_1J_2J_3(J_3-1)J_4+J_2J_3J_1(J_1-1)J_4+
J_1J_2(J_2-1)J_3J_4 \nonumber \\
&+ J_1J_2J_3J_4(J_4-1)
\Big)\nonumber \\
&+\Big(
J_1(J_1-1)(J_1-2)J_2J_3J_4 
+J_1J_2(J_2-1)(J_2-2)J_3J_4\nonumber \\
&+J_1J_2J_3(J_3-1)(J_3-2)J_4
+J_1J_2J_3J_4(J_4-1)(J_4-2)\Big)\nonumber \\
&+6J_1J_2J_3J_4
\Big]N^{-3}
=J_1J_2J_3J_4 J(J-1)(J-2)N^{-3}
\end{align}
where, similarly with the case $n=3$, 
 the first round bracket comes from 
chain diagrams of 4 beads, 
the second from 
three-leaf clover with one of the leafs  being a chain of two beads, the third from `sunflower'-type diagrams 
in which three traces are connected to 
one central trace at three separate points, and 
the last from 4-leaf clover diagrams in which all 
of 4 traces are connected at one point, respectively. 
See Fig. 3 for illustration. 
 These results precisely match to the corresponding S-matrix 
 elements  \eqref{n1amplitude}, provided the expansion 
 parameter is related by $i2/c_0^2\rightarrow 1/N$.

 The same results as above can also be obtained 
 using the exact general formulas \cite{formula}, 
 which are expressed as linear combinations 
 of the ratios of Gamma functions, such as (say $n=1, 2$)
 \begin{equation}
  \langle \Tr( \overline{Z}^J )\Tr(Z^{J})\rangle 
 ={1\over J+1}\Big(
 {\Gamma(N+J+1)\over \Gamma(N)}
 -{\Gamma(N+1)\over \Gamma(N-J)}
 \Big), \label{2ptgamma}
   \end{equation}
 \[
 \langle \Tr( \overline{Z}^J )\Tr(Z^{J_1})\Tr(Z^{J_2})\rangle 
 ={1\over J+1}
 \Big(
 {\Gamma(N+J_1+J_2+1)\over \Gamma(N)}
 +{\Gamma(N+1)\over \Gamma(N-J_1-J_2)}
 \]
 \begin{equation}
 -{\Gamma(N+J_1+1)\over \Gamma(N-J_2)}
 -{\Gamma(N+J_2+1)\over \Gamma(N-J_1)}
 \Big), \label{3ptgamma} \quad etc
 \end{equation}
for these type of correlators. See the Appendix. 
These exact formulas will also be used for a study of the 
higher-genus effect in the next section. 

The computation of higher-point cases 
and more general configurations of initial and 
final states becomes 
increasingly cumbersome. To convince ourselves 
the validity of agreement further, it is 
useful to consider the limit of large momentum $J$. 
Let us define a unitary scattering 
operator $S=\exp {\cal V}$ by 
\begin{equation}
S\alpha_{+}(\eta)S^{-1}=
\sum_{n=0}
{1\over n!}\overbrace{[{\cal V},[{\cal V},[\cdots ,
[{\cal V}}^{n\ times}
, \alpha_+(\eta)\overbrace{]\cdots ]]]}^{n\ times}
= \alpha_-(\eta).
\end{equation}
It should be kept in mind 
that here the commutators should be understood 
as Poisson brackets, since we are in the tree approximation 
where the operator ordering is irrelevant. 
In the limit of large momentum, we can replace 
the \eqref{inoutrelation} by ($\kappa=2/c_0^2$) 
\begin{equation}
\alpha_{\pm}(\eta)
=\sum_{p=1}^{\infty}
\kappa^{p-1}
{(\mp i\eta)^{p-1}\over p!}
\int d^p \xi \delta(\eta-\sum \xi_i)
\Big(\prod \alpha_{\mp}(\xi_i)\Big)
\label{largemomentum}. 
\end{equation}
Then, we find that the scattering operator can be 
written in a simple closed form as 
\begin{equation}
S=\exp\Big[i{\kappa\over 6}\int d\xi_1\int d\xi_2 
\int d\xi_3 \, \delta(\xi_1+\xi_2+\xi_3) 
\alpha_+(\xi_1)\alpha_+(\xi_2)\alpha_+(\xi_3)\Big]
\end{equation}
For example, the second term $n=1 \, (p=n+1)$ is 
\[
[i{\kappa\over 6}\int d\xi_1\int d\xi_2\, 
\alpha_+(\xi_1)\alpha_+(\xi_2)\alpha_+(-\xi_1-\xi_2), 
\alpha_+(\eta)]
={i\eta\kappa\over 2}\int d\xi_1\, 
\alpha_+(\xi_1)\alpha_+(\eta-\xi_1)\equiv O_1
\]
The third term ($n$=2) 
is equal to  
\[
{1\over 2}[i{\kappa\over 6}\int d\xi_1\int d\xi_2\, 
\alpha_+(\xi_1)\alpha_+(\xi_2)\alpha_+(-\xi_1-\xi_2), 
{i\eta\kappa\over 2}\int d\xi_1\, 
\alpha_+(\xi_1)\alpha_+(\eta-\xi_1)]
\]
\[
={i\kappa^2\eta\over 2}\int d\xi_1\int d\xi_2 \, (i\xi_1)
\alpha_+(\xi_1)\alpha_+(\xi_2)\alpha_+(\eta-\xi_1-\xi_2)
\]
\[=
{i\eta\kappa^2\over 2}
\int d\xi_1\int d\xi_2 \int d\xi_3 \, \delta(\xi_1+\xi_2+\xi_3-\eta)
\, {1\over 3}\, i(\xi_1+\xi_2+\xi_3)
\alpha_+(\xi_1)\alpha_+(\xi_2)\alpha_+(\xi_3)
\]
\[=
{(i\eta)^2\kappa^2\over 6}
\int d\xi_1\int d\xi_2 \int d\xi_3 
\delta(\xi_1+\xi_2+\xi_3-\eta)
\, 
\alpha_+(\xi_1)\alpha_+(\xi_2)\alpha_+(\xi_3)\equiv O_2
\]
To establish the general terms 
in the claimed form, we can proceed by mathematical
 induction. 
Suppose the $n$-th term in the expansion is given by 
\[
O_n\equiv {(i\eta)^{n}\kappa^{n}\over (n+1)!}
\int\cdots\int d\xi_1\cdots d\xi_{n+1}\, 
\delta(\xi_1+\cdots +\xi_{n+1}-\eta)
\alpha_+(\xi_1)\cdots \alpha_+(\xi_{n+1}).
\]
Then by similar manipulation as for $n=2$, 
we can check that the next term $n+1$ is indeed given by 
\[
{i\over (n+1)}[i{\kappa\over 6}\int d\xi_1\int d\xi_2\, 
\alpha_+(\xi_1)\alpha_+(\xi_2)\alpha_+(-\xi_1-\xi_2), 
O_n]
=
O_{n+1}.
\]

When we use the Euclidean convention from the beginning, 
we obtain the same result by making a replacement 
${\cal V}\rightarrow {\cal V}_3$. 
\begin{equation}
{\cal V}_3 ={1\over 2N}\sum (\alpha(J_1+J_2)\alpha(-J_1)\alpha(-J_2) 
+ \alpha(J_1)\alpha(J_2)\alpha(-J_1-J_2))
\label{okuyamaform}
\end{equation}
by restricting the value of (purely imaginary) momentum 
by discrete integers $J_i$'s. The vertex is 
essentially the time-integrated 
interaction Hamiltonian \eqref{h3}, 
restricted on the mass-shell,  of collective field theory 
\[
{\cal V}_3 \sim \int {d\tau\over 2\pi}\, H_3(\tau) .
\] 
In fact, it is not surprising that in the limit 
of large momentum, the intermediate states behave almost 
like on-mass-shell states. 

Interestingly, precisely the 
same formula \eqref{okuyamaform} 
has previously been proposed in \cite{okuyama}
which seems to be motivated purely on a 
combinatorial basis that this overlap-type 3-point 
vertex is simulating the structure of 
splitting and joining of matrix traces.\footnote{
One of the present author (T. Y.) thanks K. Okuyama for 
a discussion on his work. 
} 
The suggestion made in the last reference was that this 
formula should be valid for arbitrary genus 
in the BMN limit ($J, N\rightarrow 
\infty, g_2=J^2/N=$fixed).  Our results can thus be 
regarded as a generalization of this proposal for arbitrary 
finite $J$ at least in the planar approximation, 
by making clear the basis for the correspondence 
between the $c=1$ matrix model and the bubbling geometry. 
In particular our S-matrix interpretation implies that for finite  $J$ the 
above vertex would not be appropriate for evaluation of 
higher then 3-point correlators. For that one should uses the 
nontrivial c=1 string vertex as in \cite{djr}.

\section{Droplet scattering 
from the viewpoint of the $c=1$ matrix model}
\setcounter{equation}{0}

\subsection{$c=1$ scattering interpretation 
of the correlators}
Since the function $f$ does not depend 
on spacetime dimensions, we can consider a 
matrix quantum mechanics to represent the same 
correlators. 
Traditionary, we  use a complex matrix for this purpose 
with special constraint called the `lowest Landau-level 
condition',  as originally discussed in \cite{cjr} 
and followed by most literature related to 
this subject. 

     From the reduction, it is clear that one can
 start with the ordinary Hermitian matrix model
 (of Lorentzian signature), 
$
\int d\tau \, {1\over 2} \Tr\Big[
\Big({dM\over d\tau})^2 - M^2
\Big]
$, 
by considering the correlators in the following normal-order 
prescription, 
\begin{equation}
\langle 
{\rm :}{\cal O}^J_{(J'_1, \ldots, J'_{m})}(\tau')_M{\rm :}\, 
{\rm :}{\cal O}^J_{(J_1, \ldots, J_n)}(\tau)_M{\rm :}\rangle
=F(\{(J'), (J)\}, N)D_1(\tau,\tau')^J
\end{equation}
where $D_1(\tau,\tau')\propto \exp(i|\tau-\tau'|)$ and 
\begin{equation}
{\cal O}^J_{(J_1, \ldots, J_n)}(\tau_1)_M
 \equiv 
\Tr\Big(M^{J_1}\Big)\Tr\Big(M^{J_2}\Big)
\cdots \Tr\Big(M^{J_n}\Big).
\end{equation}
The normal product symbol $\np \cdots \np$ indicates that 
 no contraction is allowed inside. 
This interpretation was emphasized in ref. \cite{yone} 
 to be useful for extending discussions to 
 more general 1/2-BPS operators in which the 
 other SO(6) states than \eqref{halfbpsope} appear 
 on an equal footing. 
 
In our case, it is more natural 
 to consider the model with inverted harmonic potential 
of negative sign, 
in order to utilize 
 the standard $c=1$ matrix model to which the 
 Euclideanized bubbling geometries fit well,  
\begin{equation}
S_{c=1}=\int d\tau \, {1\over 2} \Tr\Big[
\Big({dM\over d\tau})^2 +M^2
\Big]
\end{equation}
and consider the operators 
\begin{equation}
\Pi_{\pm} =(M\pm \dot{M})/\sqrt{2}. 
\end{equation}
The correlators are then given as 
\begin{equation}\langle 
{\cal O}^J_{(+;J'_1, \ldots, J'_m)}(\tau')\, 
{\cal O}^J_{(-;J_1, \ldots, J_n)}(\tau)
\rangle
=F(\{(J'), (J)\}, N)D_1(\tau,\tau')^J
\label{c=1form}
\end{equation}
with 
\begin{equation}
{\cal O}^J_{(\pm; J_1, \ldots, J_n)}(\tau_1)
 \equiv 
\Tr\Big(\Pi_{\pm}^{J_1}\Big)\Tr\Big(\Pi_{\pm}^{J_2}\Big)
\cdots \Tr\Big(\Pi_{\pm}^{J_n}\Big) .
\end{equation}
Note that we now have $D_1(\tau, \tau')\propto \exp(-|\tau-\tau'|)$. The normal-order prescription is automatically taken into account 
since $\langle \Pi_+\Pi_+\rangle =\langle \Pi_-\Pi_-\rangle =0$. 
In the language of the complex matrix model 
of ref. \cite{cjr}, the matrix operators 
$\Pi_{\pm}$ are related to the following canonical 
decomposition 
\[
Z = {1\over{\sqrt 2}} (A^{\dagger} + B) , \quad 
\bar{Z} = {1\over{\sqrt 2}} (A + B^{\dagger} )
\]
with $A, A^{\dagger}$ being replaced by $\Pi_{\pm}$. 
The lowest Landau level condition amounts to 
eliminating the additional 
 canonical pair ($B, B^{\dagger}$). 

For evaluating the correlators, we 
can use the collective-field representation. 
Remember that in the planar approximation 
of matrix models, the collective field theory 
is essentially a phase-space representation 
of free-fermion liquid. 
For our purpose, it is most convenient to use the
coherent-state representation of the phase space 
in which $\Pi_- (\Pi_+) \sim A^{\dagger} (A)$ are 
regarded as 
generalized coordinate ($z$) and momentum ($\alpha$), 
respectively. 
\begin{align}
\Tr \left(\Pi_-^J \right) 
& \rightarrow \int {dz\over 2\pi} \int^{\alpha}\, d\alpha
\,  z^J = \int {dz\over 2\pi} \, z^J \alpha (z) = \alpha_{-J}, \cr 
\Tr \left(\Pi_+^J\right) & \rightarrow \int {dz\over 2\pi} \int^{\alpha} \, d\alpha \, \alpha^J = \int {dz\over 2\pi} \, {\alpha(z)^{J+1}\over J+ 1}.
\end{align}
It is relevant to note that there is also a (dual) coherent state representation in which
\begin{align}
\Tr \Pi_-^J & = \beta_J = \int {dz\over 2\pi} z^{-J} \beta (z), \cr
\Tr \left( \Pi_+^J\right) & = \int {dz\over 2\pi} 
{\beta (z)^{J+1}\over J+1}.
\end{align}
In the (analytically continued) scattering picture, 
the operators $\, \alpha_J\,$ or 
$\beta_J\,$ will be seen to coincide 
with $in$ ($out$) fields respectively.  
The existence of the dual representation can  be 
related to 
the freedom of two seemingly different choices 
 for the 
sign of $\mu$, appearing in the construction of the collective Hamiltonian  as has already been alluded to in section 3. 

The correlator then becomes
\[
\langle 0 \vert \int {dz_1\over 2\pi} \, 
{\alpha (z_1)^{J_{1}'+1}\over J_{1}' + 1} 
 \cdots \int {dz_m \alpha (z_m)^{J_{m}'+1}\over J_{m}' + 1 }
  \,\, \alpha_{-J_{1}} \alpha_{-J_{2}} \cdots \alpha_{-J_{n}} \vert 0 \rangle .
\]
In this picture, the operators $\, \Tr \left( \Pi_-^{J} \right)\,$ are simply creation operators while the operators $\, \Tr \left( \Pi_+^{J'}\right)\,$ are nontrivial polynomials.  
It is a theorem that we shall prove below that they are generated by the analog of the $\, (c=1)\,$ $S$-matrix operator. 
 We have
$$
\Tr \left( \Pi_+^J\right) = S^{-1} \alpha_J S.
$$
Consequently, the above matrix elements are 
$$
\langle \quad \rangle = \langle 0 \vert \alpha_{J_{1}'} 
\cdots \alpha_{J_{m}'} S \alpha_{-J_{1}} \alpha_{-J_{2}} \cdots \alpha_{-J_{n}} \vert 0 \rangle
$$

To demonstrate 
that $\, S\,$ is  nothing but 
the $S$-matrix of the  $\, c=1\,$ theory, 
we recall the analysis through collective 
field theory \cite{je}. One
considers 
\begin{equation}
\Tr \Pi_+^l\rightarrow \Tr \left( (P + X \right)/\sqrt{2})^{ik} 
= W_{ik}^{(+)}, 
\end{equation}
which is represented as 
\begin{equation}
W_{ik}^{(+)} = \int {dx\over 2\pi} \left( {\alpha_+^{ik+1}\over ik+1} - {\alpha_-^{ik+1}\over ik+1} \right).
\end{equation}
This operator has an exact time evolution, 
being an eigenstate of the Hamiltonian 
$H=\Tr(P^2-X^2)/2 =\Tr (\Pi_-\Pi_++\Pi_+\Pi_-)/2$.  
A multiplication by $\, e^{-ikt} \,$ makes
 it into a constant of motion. Evaluating
  this operator at early $\, ( t\rightarrow - \infty )\,$ 
  and late $(t\rightarrow + \infty )\,$ times gives 
  the $\, S$-matrix of the theory.  One has
$$
\alpha_{\pm} (x) = \pm x \mp {1\over x} \left( \mu \mp \hat{\alpha}_{\pm} (\tau) \right)
$$
with $\, x \sim \sqrt{\mu /2} e^{\tau}\,$.  
At early time, we will have only a left moving 
wave packet given by $\, \hat{\alpha}_- ( t + \tau ) \,$ 
and at late time we have a right moving packet 
defined by $\, \hat{\alpha}_+ (t- \tau )$.  
The evaluation of $\, W_{ik}\,$ was performed 
in detail in \cite{jrv}. At early time, 
with the left-moving packet,
$$
W_{ik} \rightarrow - 
{\left(\sqrt{2\mu}\right)^{ik+1}\over ik+1} \int 
{d\tau\over 8\pi} \, e^{-ik\tau} \sum_{p=1}^{\infty}
 {(ik+1)!\over (ik+1 - p)!p!} \left( {\alpha_-\over\mu}\right)^p.
$$
At late time (with a right moving packet) only a single mode survives in the limit giving
$$
W_{ik} \rightarrow \left( \sqrt{2\mu} \right)^{ik+1} \int {d\tau\over 8\pi} \, e^{ik\tau} \, {\alpha_+ (\tau )\over \mu}.
$$

Consequently, one obtains a relationship 
between  the outgoing modes and the incoming ones
\begin{equation}
\int d\tau e^{ik\tau} \hat{\alpha}_+ (\tau ) = - \int {d\tau\over ik+1} \, e^{-ik\tau} \left( 1 + {\hat{\alpha}_-\over \mu} \right)^{1+ik}
\end{equation}
which is equivalent with \eqref{inoutrelation} obtained 
on the basis of a droplet picture on the bulk side. 
The $\, c=1\,$ $\,S$-matrix is defined by the transformation
$
\hat{\alpha}_+ = S^{-1} \hat{\alpha}_- S.
$
The identification
$$
\hat{\alpha}_- (z) \leftrightarrow \alpha (z) , 
\hat{\alpha}_+ (z) \leftrightarrow \beta (z) .
$$
then establishes the statement that the 1/2-BPS correlators coincide
with the scattering amplitudes of the  c=1 theory. 

We emphasize that this 
holds without the so-called leg factors which appear in the physical interpretation of the 
c=1 matrix model . Also the analysis given above
was done
in the tree (or planar) approximation. 

We note that this correspondence is  related to the
observation of \cite{akkn} where an identification between finite temperature c=1
amplitudes and the normal matrix integral was described.
In contrast to the AdS/CFT interpretation
 that we were concerned with
in the present work, 
the correspondence discussed in \cite{akkn} (see also \cite{mm})
gives a special integration measure on the
complex matrix model side such that equivalence is achieved with
finite temperature correlation functions. 

\subsection{Case of higher-genus }
Thus far we have given a basis for the 
correspondence between the 1/2-BPS correlators and the 
$c=1$ S-matrix from both sides of bulk and 
boundary theories at  the planar approximation. 
It is then of interest to see to what extent the 
correspondence will be valid beyond this approximation. 
On the bulk side, evaluating string-loop 
effects in the (E)AdS background is an unsolved 
problem. On the side of the gauge theory and 
the matrix models, 
the usual fermion representation based on the positive 
harmonic potential gives a rigorous definition 
of the correlators 
for arbitrary finite $N$ and integer $J$. 
In principle, there must 
be a version of $c=1$ fermion representation with negative 
harmonic potential, which gives the 
identical function $F$ for finite $N$ and $J$. 
It would require a special regularization 
in dealing with the negative-sign harmonic potential 
by appropriately taking into account the differences 
discussed in section 2. We will not elaborate on 
such a direction in the present note, since the problem 
is rather a matter of interpretation. 

Instead, we expect on the ground of 
universality 
that the S-matrix of the  $c=1$ model defined 
by the usual 
double scaling limit 
gives the right answer for large $N$ and $J$. 
In the tree approximation, 
the large $J$ limit was not necessary, as is reasonable 
since the momentum along the R-charge direction 
must be conserved and internal momenta in the 
tree approximation are fixed by external momenta.  
It is natural to expect that the discreteness of $J$ would be washed out 
in the limit of large $J$ even for internal momenta. 
 
In the rest of this section, we present a piece 
of evidence for this expectation by studying the 
simplest nontrivial case, $1\rightarrow 1$ amplitude,
 of higher-genus effect in the leading large-$J$ approximation. 
In the coherent state representation, one has 
the Hamiltonian for a single fermion
\begin{equation}
h = \left( \hat{a}_+ \hat{a}_- + {1\over 2} \right) - \mu.
\end{equation}
The wave-function can be taken as functions of $\, a_-\,$ or (the dual picture) of $\, a_+\,$.  One has the simple transform from one basis to another
\begin{equation}
\psi_- (a_+ ) = {1\over \sqrt {2\pi}} \int da_- e^{-a_{+} a_{-}} \psi_+ (a_- ).
\end{equation}
The wave-functions both obey
\begin{equation}
{1\over 2} \left( \hat{a}_+ \hat{a}_- + \hat{a}_- \hat{a}_+ \right) \psi_{\pm} = \left( i\partial_{\tau} + \mu \right) \psi_{\pm}.
\end{equation}
In the coherent state representation with $\, \hat{a}_- = a \, \hat{a}_+ = \partial/\partial a \,$, one has the equation
\begin{equation}
\left( 
 a {\partial\over\partial a} + {1\over 2} - \mu \right) \psi_{+,k} (a) = k \psi_{+,k} (a).
\end{equation}
   From the viewpoint of the fermion phase-space $\,(x,p)\,$, 
creation-annihilation coordinates are  the null plane coordinates
$
a_{\pm} \rightarrow x_{\pm} \equiv (p \pm x)/\sqrt{2}. 
$
The discussion of scattering theory in the fermionic picture can be found in \cite{mpr,akkt}. 

The fermionic wave-functions
$
\psi_{\pm} (x_+ )
$
obey the analogous Schr\"{o}dinger eqs. 
The equation is solved by
\begin{equation}
\psi_{+,k} (x_- ) = A_k x_-^{-i(k-\mu )} \Theta (x_-) + B_k (-x_-)^{-i(k-\mu)} \Theta (-x_-)
\end{equation}
with an analogous solution for $\, \psi_-$.  
The (Gaussian) transform then leads to 
the relation between the $in$ and $out$ fermion 
wave-function with the following 
reflection coefficient (see \cite{olough})
\begin{equation}
R (k-\mu ) = {\Gamma \left({1\over 2} - i (k-\mu )\right) \over {\sqrt 2\pi}} \left[ e^{i{\pi\over 2} ({1\over 2} - (k-\mu))} + \gamma \, e^{-i{\pi\over 2} (k-\mu )} \right]
\end{equation}
where $\, \gamma\,$ specifies the boundary conditions.

In conformal field notation, the fermions have the expansion
\begin{align}
 \psi (z) & = \sum z^{-n} \psi_n , \\
\psi^+ (z) & = \sum z^{-n} \psi_n^+ 
\end{align}
with $\,  n\epsilon  Z + {1\over 2}$, and
\begin{equation}
\left\{ \psi_n , \psi_m^+ \right\} = \delta_{n+m,0}.
\end{equation}
The reflection coefficient is defined by
\begin{equation}
\psi_{-n}^{{\rm out}} = (R \psi)_{-n}^{{\rm in}}
\end{equation}
where
$
\psi^{{\rm in}}(z) \equiv \psi_-(z) , \psi^{{\rm out}} (z) \equiv \psi_+ (z)
$. 
Standard Bosonization formulas then define the corresponding in (0ut) bosonic fields.  To make contact with the previous notation we have
\begin{equation}
\psi_{\pm}^\dagger (z) \psi_{\pm} (z) = \alpha_{\pm} (z)
\end{equation}
The $\, \hat{S}$-matrix is then given by
\begin{equation}
\langle 0\vert \prod_{k=1}^m \sum_n \left( R^* \psi^{\dagger}\right)_n \left( R \psi \right)_{l_{k}+n} \, \alpha_{-j_{1}} \alpha_{-j_{2}} \cdots \alpha_{-j_{n} }\vert 0\rangle .
\end{equation}

To compare this result with the 
1/2-BPS correlators, let us use the known results of 
genus expansion of two-point amplitude given in \cite{mpr}. Using their notation,  the contributions up to 3 loops 
in the limit of large momentum $q$ 
are given as 
\begin{equation}
R(q,-q)_1  \approx - {1\over 24} 
q^5, \quad 
R (q, - q)_2  \approx {3\over 5760} q^9 , \quad 
R(q, -q)_3  \approx -{9\over 2903040}  q^{13} 
\end{equation}
where we have omitted a factor
$(\Gamma (-|q|))^2 \,\, \mu^{\vert q\vert} \ $ 
multiplying each term for brevity.  

On the other hand, the exact 2-point correlator reads 
(see \eqref{2ptgamma})
\[
G (J) = {1\over J+1} \left[
{\Gamma (N+J+1)\over \Gamma (N) } - {\Gamma (N+1)\over \Gamma (N-J )}
\right].
\]
Using the expansion formula for the ratio of Gamma functions 
as discussed in the Appendix, we find 
$$
G (J) \approx {N^{J+1}\over J+1} \sum_{n=0}^{\infty} \, {1\over N^{n}} \left(
{\beta\atop n}\right) \left[ \left( {J\over 2} \right)^n - \left(
-{J\over 2} \right)^n \right].
$$
Only the odd $\, n = 2k+1 \,$ terms are nonzero, giving the expansion
$$
N^{-J} G (J) \approx \sum_{k=0}^{\infty} \, {2\over JN^{2k}} \left(
{J\atop 2k+1} \right) \, \left( {J\over 2} \right)^{2k+1}
\approx {2N\over J}\sinh {J^2\over N}, 
$$
or
\begin{equation}
N^{-J} G(J) \approx J \left[ 1 + {1\over N^2} \, 
{J^4\over 3!2^2} +
{1\over N^4} \, {J^8\over 5! 2^4} + {1\over N^6} 
\, {J^{12}\over 7! 2^6} +
\cdots \right].
\end{equation}
The odd denominators give  coefficients 24, 1920, 322560 agreeing with the
three coefficients in the $\, c=1\,$ theory, up to 
an overall factor $i$ which should be combined to 
the $\delta$-function of energy conservation as 
in the planar cases and also to sign factor $(-1)^n$ for $n$ loops.

We can confirm that this agreement is valid 
to all orders. For this purpose, it is useful to 
recall the exact expression given in \cite{mpr} 
for the matrix model two-point function. 
\begin{equation}
{\partial\over \partial\mu}
R(q, -q) = \Gamma (-|q|)^2
\,\, {\rm Im} \left\{   e^{i\pi \vert q\vert /2}
\left( {\Gamma \left( |q| + {1\over 2} - i\mu \right)\over
 \Gamma \left({1\over 2} - i\mu \right) } - 
 {\Gamma \left({1\over 2} -
i\mu\right)\over\Gamma \left(-|q| + {1\over 2} - i\mu \right)} \right)
\right\}
\end{equation}
which gives the expansion
\begin{equation}
R (q, - q ) = (q\Gamma (-|q|))^2 \,\, \mu^{\vert q\vert} \, \left\{
{1\over\vert q\vert} - \mu^{-2} \cdot {1\over 24} 
\cdot (|q| - 1 ) (q^2 - |q|
- 1 ) + \cdots \right\}
\end{equation}
In the large $\, q\,$ limit, one notices the identity
$$
{\partial\over\partial\mu} \, R (q , - q) = \mu^{-1} \vert q \vert \, R
(q, - q )
$$
which follows from the fact that the factor $\, \mu^{\vert q\vert} $
gives the dominant effect, remembering that we keep 
only the leading large $q$ term in each 
order of $1/\mu^2$ expansion. Consequently at large $\, q \,$ (only) we
can use the formula
\begin{equation}
R (q, - q ) \approx {1\over \vert q\vert} \mu {\partial\over\partial\mu}
\, R (q, - q), 
\end{equation}
and  we have 
\[
 \Gamma(-|q|)^{-2}
 R(q, -q) \approx {1\over 2|q|i} \left\{ e^{i{\pi\over 2} \vert q \vert} \left(
{\Gamma (q + {1\over 2} - i\mu )\over \Gamma ({1\over 2} - i\mu)} -
{\Gamma ({1\over 2} - i\mu)\over\Gamma (-q + {1\over 2} - i\mu )}
\right) \right.
\]
\[\hspace{3.7cm}
 \left.- e^{-i{\pi\over 2} \vert q \vert} \left( {\Gamma (q + {1\over
2} + i\mu)\over\Gamma ({1\over 2} + i\mu )} - {\Gamma ({1\over 2} +
i\mu)\over\Gamma ( - q + {1\over 2} + i \mu )} 
\right)\right\}.
\]
Using the same expansion formula for the two terms inside 
the round bracket as before, 
one generates at large
$\, q\, $ the terms
\[
\pm {(\mu)^q\over q} \sum_{n=0}^\infty \, \left( \mp i\mu \right)^{-n} \, \left(
q\atop n \right) \left[ \left({q\over 2} \right)^n - 
\left(-{q\over 2} \right)^n
\right].
\]
Since in the sum only the odd $\, (n = 2k+1)\,$ terms contribute these
two contributions add up  giving an agreement with the 
gauge-theory correlator $G(J)$, under the replacement $-\mu^2 \rightarrow N^2$ 
with an overall factor $\pm i$.  The sign is consistent with the 
correspondence of genus-expansion parameters 
we found at the planar level. 

The fact that $|\mu|\sim N$ implies that the fermion levels 
must be filled evenly from the 
top of the potential for the ground state, quite differently from the usual 
double scaling limit in defining higher genus contributions 
in the traditional treatment of the $c=1$ matrix model. 
This of course reflect the feature emphasized in the 
beginning of this subsection. 

\section{Conclusion}
We have argued that the two-point 
correlators of 1/2-BPS multi-trace operators are 
interpreted as the S-matrix of the $c=1$ Hermitian 
matrix model. 
The correspondence of both sides are 
fairly tight in the planar approximation. However, 
as we have cautioned, the situation is somewhat 
different if we consider the non-planar case, where 
we have presented evidence for the correspondence 
only in the limit of large momentum $J\rightarrow \infty$. 
To establish exact correspondence for 
finite $J$ and large but finite $N$, we need a modified 
definition of the $c=1$ model, in regard to the 
level spacings in filling fermion states 
 from the top of the potential, not from the bottom, 
 as we have alluded to in section 2. 
It must be 
defined such that 
 the `ground state' of the model 
becomes equivalent with that of the matrix model 
with positive harmonic potential and consequently 
we can have a one-to-one 
mapping even for
 excited states (again from top to down) to 
those of the positive harmonic potential.  

We may also hope that, in the large $J$ limit, we 
can discuss the genus corrections from the viewpoint of 
pp-wave string field theory. However, at the present state 
of development, even the question of what is the 
right string-field vertex for this purpose is not settled. 
For some recent works related to this question, 
see \cite{semenoffetal}.

We have emphasized that unlike the known 
correspondence between $c=1$ matrix model and the 
S-matrix  of 2D string theory, 
the leg factor is not necessary. In ref. \cite{dy}, 
it has been established that we need a special leg-like factor 
in order to relate the 3-point correlator of the BMN operator 
with the Euclidean S-matrix in the AdS bulk from boundary to boundary. 
However, for the extremal correlators 
such as 
\[
\langle \Tr (\bar{Z}^{J_1+J_2})(x)\Tr (Z^{J_1})(y)\Tr (Z^{J_2})(y')\rangle ,
\] the corresponding 3-point (on-shell) vertices in the bulk vanish and 
simultaneously the leg factor diverges, so that the product 
of 3-point vertices and the leg factor gives finite 
correlators. In the limit $y\rightarrow y'$, these extremal 
3-point correlators reduce smoothly to two-point correlators 
which are dealt with in the present work. Similar arguments 
apply also to higher-point extremal correlators. 
It is quite significant that 
the droplet picture provides us a direct correspondence 
for the correlators and Euclidean S-matrix in the bulk 
without such complications. It would be very 
interesting if there is any extension of our results to 
non-BPS operators. 

Indeed, the c=1 interpretation might be of relevance to the
full AdS string where analogous structures and $c=1$ type 
scaling \cite {gh,ms} 
has been observed recently. The latter correspondence regarding the 
{\it world-sheet} S-matrix 
points towards capturing the mixing among excited but {\it single-body} 
string states. In contrast to this, in our case,  the $c=1$ matrix model 
is relevant in understanding the {\it multi-body}  
states of strings, restricted to the ground state of spin chain states.

We  noted parallels with the relation 
between the so-called normal (complex) matrix model 
and the $c=1$ model at finite temperature \cite{akkn}.
It would be of interest 
to clarify the connection further. 

Finally, we mention also the recent investigation of \cite{bdrt}
which discusses correlators and topology changes in the
1/2 BPS sector.

\vspace{1cm}
\noindent
Acknowledgements

One (T.Y.) of the present authors thanks the 
Physics Department of Brown 
University and the Center for Theoretical Physics, MIT 
for hospitality. Main part of the present work was 
done during T.Y.'s visit to both institutions. 

The work of A. J. is supported in part by the Department of Energy under Contract
DE-FG02-91ER40688,Task A .

The work of T. Y. is supported in part by Grant-in-Aid for Scientific Research (No. 13135205 (Priority Areas) and No. 16340067 (B))  from the Ministry of  Education, Science and Culture, and also by Japan-US Bilateral Joint Research Projects  from JSPS.

\appendix 
\section{Genus expansion from exact formulas}
\setcounter{equation}{0}
\renewcommand{\theequation}{\Alph{section}.\arabic{equation}}
\renewcommand{\thesubsection}{\Alph{section}.\arabic{subsection}}
Here we present some formulas which are useful for studying 
 genus expansion on the basis of 
exact formulas for correlators. First we need an expansion
 formula \cite{gammaformula} for 
the ratio of Gamma functions: 
\begin{equation}
{\Gamma(z+\alpha)\over \Gamma(z+\beta)}=
\sum_{n=0}^{\infty}{\Gamma(\alpha-\beta +1)\over 
n! \Gamma(\alpha-\beta -n+1)}
B_n^{(\alpha-\beta+1)}(\alpha)z^{\alpha-\beta -n}, 
\label{expansion}
\end{equation}
where the coefficient function $B_{n}^{(\mu)}(x)$  
is the generalized Bernoulli polynomial \cite{genebernoulli} expressed as
\[
B_n^{(\mu)}(x)
=\sum_{k=0}^n {k!\over (2k)!} 
\left({n \atop k}\right)\left({\mu + k-1 
\atop k }\right)\sum_{j=0}^k (-1)^j
\left({k \atop j}\right)j^{2k}(x+j)^{n-k}
\]
\begin{equation}\times 
F[k-n, k-\mu; 2k+1; {j\over x+j}]. 
\end{equation}
 
In our case, the explicit expressions of the coefficients are 
\begin{align}
B_0^{(\alpha-\beta +1))}(\alpha)&=1, \cr
B_1^{(\alpha-\beta +1)}(\alpha)&={1\over 2}(\alpha +\beta -1) , \\
B_2^{(\alpha-\beta +1)}(\alpha)&=
{1\over 4}(\alpha + \beta)^2 -{7\over 12}\alpha -{5\over 12}\beta 
+{1\over 6} , \\
B_3^{(\alpha-\beta +1)}(\alpha)&=
{1\over 8}(\alpha +\beta)^3 -{1\over 2}
(\alpha +{\beta \over 2})(\alpha +\beta) +{3\over 8}\alpha 
+{1 \over 8}\beta. 
\end{align}
In particular for  $\alpha -\beta\gg 1$, we have 
an asymptotic form
\begin{equation}
{\Gamma(z+\alpha)\over \Gamma(z+\beta)}\sim 
\sum_{n=0}^{\infty}\left({\alpha-\beta \atop 
n}\right){(\alpha + \beta)^n \over 2^n} z^{\alpha-\beta -n}, 
\end{equation}
which can easily be understood by a `dilute-gas' approximation 
for the power-series expansion of the above ratio in $1/z$. 
Note that in this limit $(\alpha+\beta)/2$ is the average value of the 
coefficients of $1/z$ in the product 
$\prod_{n=\beta}^{\alpha-1} (1+\frac{ n}{z})$.  
This asymptotic formula is used in section 5. 

Using the exact formulas \cite{formula} such as \eqref{2ptgamma}, 
\eqref{3ptgamma} and, for $n=3$, 
\[
\langle 
\Tr(\bar{Z}^J) \Tr(Z^{J_1})\Tr(Z^{J_2})\Tr(Z^{J_3})\rangle
={1\over J+1}
\Big(
{\Gamma(N+J_1+J_2 +J_3+1)\over \Gamma(N)}
-{\Gamma(N+J_2 +J_3 +1)\over \Gamma(N-J_1)}
\]
\[
-{\Gamma(N+J_1 +J_3 +1)\over \Gamma(N-J_2)}
-{\Gamma(N+J_1 +J_2 +1)\over \Gamma(N-J_3)}
+{\Gamma(N+J_1 +1)\over \Gamma(N-J_2-J_3)}
\]
\begin{equation}
+{\Gamma(N+J_2 +1)\over \Gamma(N-J_1-J_3)}
+{\Gamma(N+J_3 +1)\over \Gamma(N-J_1-J_2)}
-{\Gamma(N +1)\over \Gamma(N-J_1-J_2-J_3)}\Big) \quad etc, 
\end{equation}
together with the above expansion formula \eqref{expansion}, 
we can check the results of section 4, although 
the algebra becomes increasingly tedious  for larger $n$.

\end{document}